\documentclass{sf2a-conf2015}
\usepackage{graphicx}
\usepackage{hyperref}
\usepackage[]{natbib}  
\usepackage{epstopdf}

\def\BibTeX{{\rm B\kern-.05em{\sc i\kern-.025em b}\kern-.08em
    T\kern-.1667em\lower.7ex\hbox{E}\kern-.125emX}}
\bibpunct{(}{)}{;}{a}{}{,}  

\begin{document}

\TitreGlobal{SF2A 2015}


\title{News from the CFHT/ESPaDOnS spectropolarimeter}

\runningtitle{ESPaDOnS news}

\author{C. Moutou}\address{CFHT, Kamuela HI 96743, USA}

\author{L. Malo$^1$}

\author{N. Manset$^1$}

\author{L. Selliez-Vandernotte$^1$}

\author{M.-E. Desrochers$^1$}

\setcounter{page}{237}


\maketitle


\begin{abstract}
The ESPaDOnS spectropolarimeter has been in use on the
Canada-France-Hawaii Telescope (CFHT) since 2004, for studying stars,
galactic objects and planets. ESPaDOnS is used in queued service
observing mode
since 2008, which
allows an optimization of the science outcome. In this article, we
summarize the new functionalities and analyses made on ESPaDOnS
operations and data for the present and future users. These
modifications include: signal-to-noise ratio based observing,  radial
velocity nightly drifts, the OPERA pipeline under development, the
measurement of H2O content in the Maunakea sky, and the use of
ESPaDOnS with the neighbour telescope Gemini.  
\end{abstract}



\section{Introduction}
The ESPaDOnS spectropolarimeter is used in three different modes:
Star-only (single fiber on the star, highest resolving power, 81000), Star+Sky (68000 resolving
power, second fiber observes the sky background), and Polarimetric
(68000 resolving power, linear or circular polarisation). With its
twin instrument NARVAL at Telescope Bernard Lyot, ESPaDOnS has the
unique capability of measuring the magnetic topologies of
stars. Recent examples of ESPaDOnS results are given in, e.g., 
\citet{2015MNRAS.453.3706D,2015MNRAS.454L...1S, 2015ASPC..496..411M,2015A&A...575A..34M,2015MNRAS.446.1988S,2015arXiv150906818W}. While ESPaDOnS is highly efficient, cover the totality of
the visible bandpass (370-1050nm) and benefits from the exquisite
conditions at Maunakea 4200m-high Observatory, its performances and
global science outcome can still be optimized in a number of ways. In
the following, we summarize the directions taken by the observatory
staff to improve the consistency, reliability and performance of
ESPaDOnS operations. This report is also meant to help present and
future users to understand their data better and/or to plan more
efficiently their future ESPaDOnS observing campaigns.
 
\section{Signal-to-noise based observations}
ESPaDoNS is operated in queued service observing mode since 2008, from Phase 2
information provided by the users: instrumental configuration,
exposure time per exposure, sequence strategy, science priorities,
finding charts, sky constraints and potentially timing
parameters. Then the queue for a given night is built from the entire
database of observations, by a queue coordinator who minimizes the
telescope motion, the instrumental changes (only two modes out of the
three can be used in a given night), and maximizes the science
outcome, while trying to accomodate all timing constraints. Backup
queues are also prepared with alternate targets. This queue
can be executed as such during the night, if the
conditions are good and stable; the observer can also pick any
observation from the queue, skipping or re-ordering observations as
needed . If conditions are bad and/or variable,
the remote observer can repeat high priority observations or switch to
low priority snapshot programs. However, when conditions are better
than median, the stellar signal increases rapidly and it
would be possible to fit in more observations and increase the overall
quality of the collected data. For this reason, the CFHT observatory
aims at offering an operational strategy based on the measured signal-to-noise
ratio (SNR) rather than fixed exposure times for ESPaDonS. Note that this
mode is already used for some MegaPrime imaging surveys \citep{2014ASPC..485...81C}, allowing an
homogeneous depths in a wide field of view. For ESPaDOnS, in addition
to adapt exposure times as a function of observing conditions, an
SNR-based operation would also allow to 
homogeneize the quality of the numerous polarimetric sequences (10 to 15 identical
observations spread over two weeks). 

In order to set up the SNR mode, the observatory needs additional
information from the users, namely: 1) the magnitude and effective
temperature of the target, 2) the goal SNR per pixel and per intensity
spectrum. By using this information, when available, we were able to
estimate that 25 to 30\% of the exposure time would have been saved, by
observing at the requested SNR and not beyond. The science validation
of this new mode requires that the users provide the observatory staff
with relevant and complete information. One way to verify this in a
systematic way is to use the instrument model (exposure time 
calculator\footnote{\url{http://etc.cfht.hawaii.edu/esp/}}) with
adequate ranges of image quality and airmass. The user can now iterate
with the different values of exposure time, external constraints and
goal SNR, in the Phase 2 web tool, until all information is consistent in the database. 

Second, the remote observer needs a realtime estimate of the SNR during
the night, or even during the exposure. This is possible with the
exposuremeter, a photodiode that collects a small part of the light
going through the spectrograph. By comparing exposuremeter counts and
final SNR as measured by the pipeline (examples in Figure 1), we were able to calibrate the
realtime flux and estimate the realtime behaviour of the SNR. This
calibration depends on the instrumental mode and requires
adjustements for the spectrograph temperature, which affects the
mechanical structure of the optical mounts. The realtime estimate of
the SNR is presently (Sept 2015) in testing, for further validation. When
deemed a robust estimate, the observer will be able to trigger the
anticipated readout of the exposure in the cases where the goal SNR
has been reached before the expected end of the exposure.  A
final complication comes from the fact that the SNR for ESPaDOnS can
be defined at several places in the spectrum, which is freely specified by
the user, while the calibration of the exposuremeter is done at a
specific wavelength (755nm, center of order 30). So it is first necessary to
use the exposure time calculator from the database to derive the goal
SNR at order 30, from the goal SNR at any wavelength provided by the user.
Users are also encouraged to use order 30 as their new reference for goal SNR
requests!

With a simultaneous monitoring of the flux count entering the
spectrograph, one could also prevent saturated exposures -for 
program relying on extreme SNR values- but mostly, the gain in
exposure time would be beneficial to the whole user community. The
SNR operation mode with ESPaDOnS could be made available in 2016A,
provided that all tools are validated and robust, and at the
discretion of the users. Indeed, some programs may be inadequate for
SNR mode observations: time critical sequences, targets of very
variable magnitude, moving targets or very faint objects.

\section{Radial velocity shifts and precision}
ESPaDOnS is not an optimized velocimeter: the spectrograph is not
actively controlled in temperature nor pressure, the fiber injection
is not scrambled, and there is no possible simultaneous wavelength
calibration as in high-precision radial-velocity
spectrographs. However, the spectral range and resolution allow to
obtain a good precision on the spectrum of Earth atmosphere bands, and
to correct at first order for the wide velocity shift that occurs during the
night. The wavelength calibration of the spectrograph, done in the
afternoon or the morning, results in an error of about 200 m/s. There
are ways to reduce this error, by improving the pipeline (see section 5).
By observing known exoplanets, previous studies have shown that
a residual scatter of $\sim$20 m/s remains on the stellar radial velocity
measurement, when the Least Square Deconvolution intensity profile is
used \citep{1997AA...326.1135D} and when the measurement is corrected for the
spectrograph shift as measured on the telluric spectrum
\citep{2007A&A...473..651M}. On a very short timescale of 1 hour, the scatter is of the
order of 10 m/s. 

The typical shift during a night is of the order of 200 to 500 m/s,
depending on the temperature and pressure evolution. The behaviour of
the drift also depends on the range of these parameters, and can be linear
or sinusoid-like, or anything in between! In the coming month, this
drift evolution will be continuously monitored on ESPaDOnS data, to
spot any change of the behaviour due to the recent installation of a
second external, passive,
thermal enclosure. Although passive, the new enclosure should limit the temperature
changes on a given timescale; also, it immediately resulted in a new
temperature range (the spectrograph has become warmer), which may
affect the mechanical distortions and the residual spectral
shifts.

\section{Water content above Maunakea}
With its spectral coverage of 370-1050nm, the ESPaDOnS spectra contain
a wealth of information about the Earth atmosphere features. As the
mean telluric band is extracted from the spectra by the
pipeline, for the correction of the spectral shift, this profile can
also be  used for calibrating the water content in the sky at the time of the observation.
The equivalent width of the telluric profile ($EW$) has been correlated to
the simultaneous measurement of the water absoption by the Caltech
Submm Observatory ontop
Maunakea\footnote{\url{http://cso.caltech.edu/tau/}}. The
following equations are used to derive the precipitable water vapor
from the ESPaDOnS spectra:
\begin{equation}
PWVesp(zenith)   = 2.5335 - 1.608 \times EW + 0.200 \times EW^2 
\end{equation}
\begin{equation}
PWVesp = 1./airmass*(PWVesp(zenith) - 0.296 \times (airmass-1.))
\end{equation}
where $PWVesp$ is the value at the observed $airmass$ and $PWVesp(zenith)$
the value at zenith.

Using more than 1 year of data where this measurement is available, we
could derive some statistical values, which are useful for specific
programs limited by the water content in the spectra and more
generally, for the future CFHT/SPIRou programs (where water content
will directly affect the radial-velocity precision).
During 50\% of the time, the PWV in the atmosphere is less than 2.5mm
H2O during nights where ESPaDOnS is used, and at the real observed airmasses.

\section{OPERA pipeline}
The standard pipeline of ESPaDOnS is Libre-Esprit \citep{1997AA...326.1135D}. It
conducts the pre-processing of the spectra, recognizes the order from
flatfield calibration exposures, performs the wavelength calibration,
uses optimal extraction to derive the final spectra, calculates
polarization from 4 polarimetric sub-exposures, and subtracts the sky spectrum
for data taken in the Star+Sky mode. Libre-Esprit, 
is embedded in the data flow software Upena and used daily to process
the ESPaDOnS nights and provide users with final products. 

The OPERA pipeline has started development a few years ago to fulfill
additional needs from the users: get intermediate products, and use a
two-amplifier readout mode of the detector that would reduce the
readout overheads by about a factor of two. OPERA is also used for the reduction of GRACES data
(see below), for which the processing parameters are different. Fine
tuning of OPERA algorithms is still underway but made a lot of
progress in the last year (Martioli, Malo et al, in prep.). Both OPERA and
Libre-Esprit products will be distributed to users in 2016, in order
for the observatory to get feedback before OPERA pipeline eventually
becomes the new standard.

Special care has been put into the spectral calibration into
OPERA. The selection of Thorium and Neon lines from the atlas and
from the observed lamp exposures has been optimized, with the result
that the radial-velocity scatter per order is of the order of 50 to 70
m/s. A combination of long and short exposures on the spectral lamp
also prevents some of the saturated lines to blur out the neighbour
orders. This gain of radial-velocity precision is promising and may
open new science capability.

Finally, the 2-amplifier mode is still being tested, but its reduction
poses no issue with the OPERA pipeline, although the noise properties
are different than in 1-amplifier images and changing along the chip. Science validation is yet needed
before this 2-amplifier mode is offered for observations.

\section{Meeting the giants: GRACES}
GRACES stands for Gemini Remote Access to CFHT ESPaDOnS
Spectrograph. It is literally the links between the neighbour 8m
telescope located at about 200m from the CFHT dome and the ESPaDOnS
spectrograph. A 270m fiber has been deployed that feeds ESPaDOnS when
it is not used by CFHT. GRACES has two spectroscopic modes (no
polarimetric mode) with resolving powers of 67k and 40k and covers the
same spectral range than ESPaDOnS, although the bluest part of the
spectrum is less efficient. Lamp calibration and fiber injection are provided by the Gemini
North GMOS instrument. First tests and science results obtained with
GRACES are described in \citet{2014SPIE.9151E..47C} and Malo et al (in prep).

As foreseen in the agreement between both observatories, the CFHT
community will receive a few nights of observing at Gemini North or
South in compensation for the use of ESPaDOnS (with a ratio 3/20). A first call for
proposals should appear in 2016B (March 2016).







\begin{figure}[ht!]
 \centering
 \includegraphics[width=0.8\textwidth,clip]{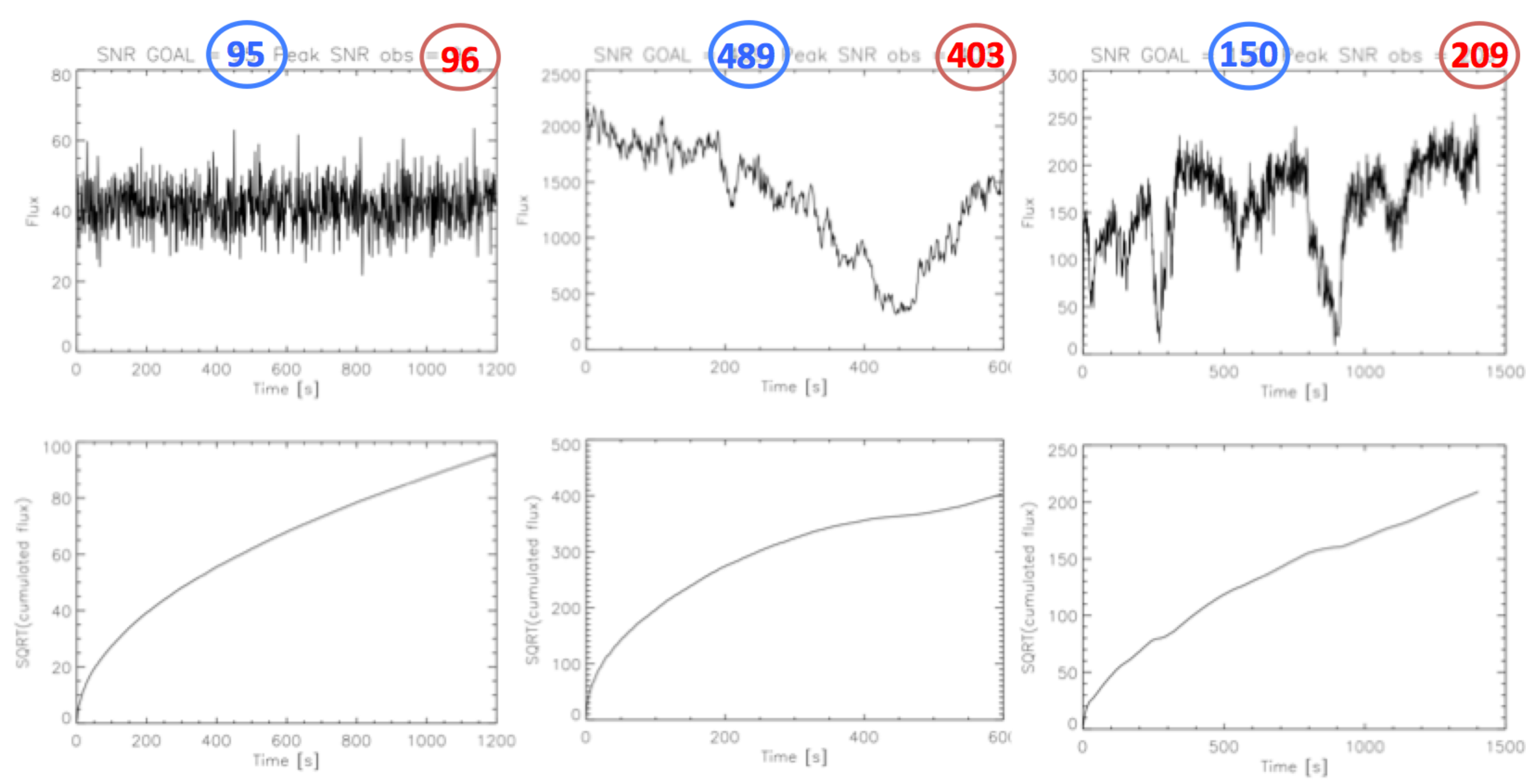}      
  \caption{Three examples of exposuremeter plots are shown: (left),
    the conditions are stable and the goal SNR (95) is achieved (96);
    (middle), the flux is altered by a cloud -or degrading image
    quality- and the goal SNR (489) is not achived (403), meaning we
    should repeat the exposure; (right) the flux is altered during the
  exposure, but the goal SNR (150) has nevertheless been exceeded, so
  that the exposure could have been reduced in realtime. The plots
  below show the realtime estimate of the SNR based on the
  exposuremeter counts.}
  \label{author1:fig1}
\end{figure}

\bibliographystyle{aa}  
\bibliography{moutou1} 

\end{document}